\documentclass[aps,prb,reprint,superscriptaddress,amsmath,amssymb,groupedaddress]{revtex4-1}

\usepackage{graphicx}
\usepackage[usenames]{color}
\usepackage{verbatim}

\begin{document}


\title{Photon drag of superconducting fluctuations in 2D systems}

\author{M.~V.~Boev}
\affiliation{Novosibirsk State Technical University, Novosibirsk 630073, Russia}

\date{\today}

\begin{abstract}
 The theory of photon drag of superconducting fluctuations in the two-dimensional electron gas is developed. It is shown that the frequency dependence of the induced current  is qualitatively similar to the case of photon drag of conventional two-dimensional degenerate electron gas. With the decreasing temperature the magnitude of the effect increases dramatically and the current of superconducting fluctuations carries an additional power of reduced temperature in comparison with the Aslamazov-Larkin contribution. The magnitude of the developed effect is expected sufficient to be visible against the conventional photocurrent background.
\end{abstract}

\maketitle

\section{Introduction}

Light absorption by a condensed matter is accompanied by the transfer of the momentum of photons, besides their energy, to  charged excitations. Such transfer results in the occurrence of the electric current in the system and this effect was called photon drag.
The magnitude and direction of photocurrent depend on many factors, such as light polarization,  incident angle and frequency. Moreover, the transport features of carriers in a condensed matter and the microscopic mechanism of light-carriers interaction drastically influence on the photon-drag. That is the reason why this effect is widely used in the study of numerous systems: semiconductors \cite{Danishevskii1970,Gibson1970}, two-dimensional electron gas \cite{Graf2000,Wieck1990,Shalygin2006}, graphene \cite{Glazov2014}, topological insulators \cite{Plank2016}, metal-semiconductor
nanocomposite \cite{Mikheev2018}, two-dimensional exciton gas \cite{Kovalev2018,Kovalev2018A,Boev2018} and metal films \cite{Vengurlekar2005,Strait2019}.

At the same time, the investigation of superconductivity phenomenon in two-dimensional systems takes a great part in condensed matter physics. Starting from thin metallic films, the samples fabrication technologies and experimental tools become suitable for the study of highly crystalline superconductors possessing extremely small thicknesses down to a monolayer \cite{Saito2016}. Among atomically thin superconductors, the systems based on transition metal dichalcogenides (TMD), e.g. MoS$_2$, have aroused interest in recent years \cite{Lu2015,Costanzo2016,Saito2016Nat,Piatti2018,Sharma2018}. The remarkable feature of such systems is the use of ionic liquid gate for creating a large density of electrons, reaching the values up to $3\cdot 10^{14} cm^{-2}$.

To date, the transition of
TMD and main-group metal dichalcogenide flakes from the normal (resistive) to superconductive phase have been studied in experiments \cite{Ye2012,Jo2015,Shi2015,Lu2018,Zeng2018,Xi2016,Zhu2018,Kang2018,Song2019}. However,  in the range of temperature close to the phase transition, $T\gtrsim T_c$,  the behaviour of TMD flakes in the electromagnetic (EM) field has not been completely studied experimentally, as well as theoretically. In this direction, it was observed that the superconductive fluctuations in the normal phase make the effect of magnetochiral anisotropy be noticeably more distinct \cite{Wakatsuki2017}.

By now, the investigations of fluctuation phenomena in superconductors have opened comprehensive facilities in  identification of fundamental properties of superconductors  \cite{Varlamov2018}. In the present work, we suggest the photon drag effect as an additional approach to the investigations of transport features of 2D superconductors in the fluctuating regime.
We explore the classical limit of this effect. It means that no transition between subbands happens. In other words, it is supposed that the incident EM-wave frequency is much less than any energy gap in the systems. In this case, the physical mechanism of photon drag just consists in the momentum transfer from the EM-wave to fluctuations.

To develop the theory, the Ginsburg-Landau (GL) approach is used \cite{Larkin2005}. Although the microscopic treatment is more exact, it is simultaneously more difficult and cumbersome than the GL one. Since our aim is to achieve a qualitative picture, the GL theory seems to be appropriate as a good approximation.

\section{Model}

\begin{figure}[t]
 \includegraphics[width=0.9\linewidth]{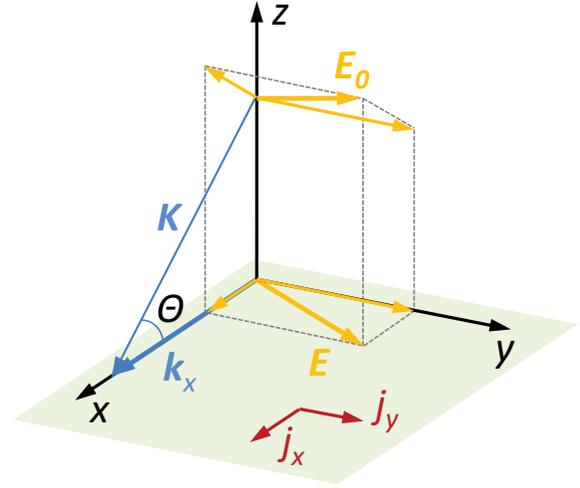}
 \caption{\label{fig0} System sketch. The EM-wave illuminate the 2D superconductor, which lies in $(x,y)$-plane, at some angle $\Theta$ and produces the electric current in $x$ and $y$ directions. Here $K$ is the wave-vector of the EM-wave, $E_0$ is the electric field amplitude, while $E$ is its projection to the $(x,y)$-plane.}
\end{figure}
Let us consider the 2D-superconductor being in the normal phase and irradiated by an electromagnetic wave with an electric field amplitude ${\bf E}_0$, wave vector ${\bf K}$ and frequency $\Omega$ (Fig.\ref{fig0}). In present work we consider a purely 2D system. Thus, the electron motion in the z-direction is neglected and  the superconducting fluctuations respond to the projection of EM-wave amplitude ${\bf E}$ on a superconductor surface only. For later computations, it is convenient to define ${\bf E}$ in the complex form:
\begin{gather}\label{EMwave1}
    {\bf E}({\bf R},t) =
    {\bf E} e^{i({\bf kR}-\Omega t)} + {\bf E}^* e^{-i({\bf kR}-\Omega t)},
\end{gather}
where ${\bf E}$ and ${\bf E}^{*}$ are complex amplitudes of electromagnetic wave and ${\bf k}$ is a projection of ${\bf K}$ to the superconductor plane. We focus on the stationary and homogenous part of the electric current, which does not vanish after the averaging in space and time.
Thus, in the lowest order of the wave amplitude, the photon drag current corresponds to the second-order response:
\begin{gather}\label{curdef}
    j_{\alpha} =
    \sigma_{\alpha\beta\gamma}({\bf k},\Omega) E_\beta E^*_\gamma,
\end{gather}
because any odd term of expansion will give zero contribution after averaging.
In Eq.\eqref{curdef} $\sigma$ is the second-order conductivity and the subscripts denote components in the Cartesian axes.
For convenience, let ${\bf k}$ to be oriented along the x-axis (Fig.\ref{fig0}). Then the system is symmetric under reflection $y\rightarrow -y$ and, therefore, only $\sigma_{xxx},\sigma_{xyy},\sigma_{yxy},\sigma_{yyx}$ are nonzero. After separating $\sigma$ to symmetric and antisymmetric parts in accordance with $\sigma_{\alpha\beta\gamma} = \sigma^s_{\alpha\gamma\beta} - \sigma^a_{\alpha\gamma\beta}$,  the current reads
\begin{gather}
    \nonumber
    \left(  {j_x
            \atop
            j_y}
    \right) =
    \\
    \left(  {\sigma^s_{xxx}|E_x|^2 + \sigma^s_{xyy}|E_y|^2
            \atop
            \sigma^s_{yxy}(E_xE_y^*+E_x^*E_y) - i\sigma^a_{yxy}(E_xE_y^*-E_x^*E_y)}
    \right),\label{curmain}
\end{gather}
where the imaginary unit in the second line is introduced to make $\sigma^a_{yxy}$ real.

Eqs. \eqref{curmain} are just a general form of the second-order response. However, the explicit expressions for conductivity are the goal.
To succeed in it, we start from the definition of an electric current in the form of the variational derivative  (further we will omit variables $({\bf R},t)$ for brevity):
\begin{gather}\label{curdef2}
    {\bf j} =
    -\frac{\delta{\mathcal F}[\Psi]}{\delta {\bf A}},
\end{gather}
where the GL free energy has the form \cite{Larkin2005}
\begin{gather}\label{GLfEnergy}
    {\mathcal F}[\Psi] =
    \alpha T_c \int dV\,    \left\{
                                \epsilon|\Psi|^2 + \xi^2|(\hat{{\bf p}}^2-2e{\bf A})\Psi|^2
                            \right\},
\end{gather}
$\Psi$ is the order parameter, ${\bf A}$ is the EM-wave vector potential, $\alpha=(4mT_c\xi^2)^{-1}$ is the GL expansion coefficient, $m$ is the electron mass, $T_c$ is the temperature of transition to the superconductive state, $\xi$ is the coherent length, $\hat{{\bf p}}=-i\nabla$, $\epsilon=\ln(T/T_c)\approx(T-T_c)/T_c$ is the reduced temperature.
In writing \eqref{GLfEnergy} it is supposed that the EM-field does not change the coefficients in the GL free energy expansion and is just included via the minimal coupling $-i\nabla \rightarrow -i\nabla -2e{\bf A}$.
Combining \eqref{curdef2} and \eqref{GLfEnergy} we can see that, as usual, the current proves to be a sum of dia- and paramagnetic terms:
\begin{subequations}\label{curterms}
    \begin{eqnarray}
        {\bf j}^D = -8e^2 \alpha T_c \xi^2 \textbf{A}|\Psi|^2,
            \label{DiaCur}
        \\
        {\bf j}^P = 4e\alpha T_c \xi^2 \textrm{Re}[\Psi^* \hat{\textbf{p}} \Psi].
            \label{ParaCur}
    \end{eqnarray}
\end{subequations}
In \eqref{curterms} the order parameter is still undefined. To proceed, let us note that including the vector potential to the GL free energy makes the order parameter dependent on it, $\Psi=\Psi({\bf A})$. To obtain the explicit expression of this dependence, we explore the Time-Dependent Ginzburg-Landau  (TDGL) equation \cite{Larkin2005}:
\begin{gather}\label{TGLe}
 \left\{
    \gamma \frac{\partial}{\partial t} + \alpha T_c \left[\epsilon+\xi^2\left(\hat{\textbf{p}}-2e\textbf{A}(r,t)\right)^2\right]
 \right\}\Psi(r,t) =
 f(r,t),
\end{gather}
where parameter $\gamma$ has both real and imaginary parts, $\gamma = \gamma'+i\gamma''$. The real part is proportional to the lifetime of fluctuating Cooper pair, $\gamma'=\alpha T_c\epsilon\tau_{GL}$ \cite{Larkin2005}. The lifetime $\tau_{GL}$ goes to infinity near the critical point and, in the BCS theory, it has the form $\tau_{GL}=\pi/8(T-T_c)$. Thus, $\gamma'=\pi\alpha/8$.  The appearance of the imaginary term, $\gamma''=-\frac{\alpha T_c}{2}\partial\ln(T_c)/\partial E_F$, in the TDGL is shown to be a consequence of the gauge invariance of GL-theory \cite{Aronov1995}. From the microscopic point of view, the origin of $\gamma''$ can arise from either the electron-hole asymmetry   \cite{Michaeli2012} or the topological structure of the Fermi surface \cite{Angilella2003}. The quantity $\gamma''$ plays the crucial role in some effect, for instance, in the fluctuation Hall conductivity \cite{Varlamov2018}. 
Further, it is assumed that $\gamma''/\gamma'\ll1$.
In eq. \eqref{TGLe}, $f$ is a Langevin random force, which defines the white noise in the system and is completely uncorrelated:
\begin{gather}\label{Langevin}
 \langle f^*(r,t) f(r',t') \rangle = 2T \gamma' \delta(r-r')\delta(t-t').
\end{gather}
Here the angle brackets designation $\langle ... \rangle$ means fluctuations averaging.
In writing the TDGL equation, we choose the gauge of EM-wave with zero scalar potential that means the connection ${\bf E} = -\partial_t {\bf A}$.
Assuming the vector potential to be a perturbation, let us utilize the method of progressive approximation, i.e. we will find the solution of \eqref{TGLe} in the form of expansion in the powers of $A$:
\begin{gather}\label{Prograpp}
 \Psi = \Psi_{0} + \Psi_{1} + \Psi_{2}...
\end{gather}
where $\Psi_{i} \sim A^i$. Since the second order response is needed, we should keep the terms $\sim A^2$ after the substitution of expansion \eqref{Prograpp} to  \eqref{curterms} yielding:
\begin{subequations}\label{curterms2}
    \begin{eqnarray}
        \langle{\bf j}^D\rangle \approx -8e^2 \alpha T_c \xi^2 \textbf{A}(\langle\Psi_{0}^*\Psi_{1}\rangle+\langle\Psi_{1}^*\Psi_{0}\rangle),
            \label{DiaCur2}
        \\\nonumber
        \langle{\bf j}^P\rangle \approx 4e\alpha T_c \xi^2 \textrm{Re}[\langle\Psi_{0}^* \hat{\textbf{p}} \Psi_{2}\rangle +\langle\Psi_{1}^* \hat{\textbf{p}} \Psi_{1}\rangle +
        \\
         + \langle\Psi_{2}^* \hat{\textbf{p}} \Psi_{0}\rangle].
            \label{ParaCur2}
    \end{eqnarray}
\end{subequations}

For the next step the explicit form of approximate solution is required.
To derive it, we rewrite \eqref{TGLe} in terms of operators:
\begin{gather}\label{TGLeOperators}
 \left\{\hat{L}^{-1} - \hat{M}_1 - \hat{M}_2
 \right\}\Psi({\bf R},t) =
 f({\bf R},t),
\end{gather}
where
\begin{subequations}\label{Operators}
    \begin{eqnarray}
        \hat{L}^{-1} = \gamma \frac{\partial}{\partial t} + \alpha T_c [\epsilon+\xi^2\textbf{p}^2],
        \label{LOperator}
        \\
        \hat{M}_1 = \alpha T_c \xi^2 2e(\hat{\textbf{p}}\textbf{A}+\textbf{A}\hat{\textbf{p}}),
        \label{M1Operator}
        \\
        \hat{M}_2 = -\alpha T_c \xi^2 (2e)^2\textbf{A}^2.
        \label{M2Operator}
    \end{eqnarray}
\end{subequations}
Thus, the formal solution of \eqref{TGLeOperators} can be obtained with multiplying \eqref{TGLeOperators} by $\hat{L}$ from the left. So, we find the following expressions for the terms in expansion \eqref{Prograpp}:
\begin{subequations}\label{Psiformal}
    \begin{eqnarray}
        \Psi_0 ({\bf R},t) = \hat{L} f({\bf R},t),
        \label{Psi0formal}
        \\
        \Psi_1 ({\bf R},t) = \hat{L} \hat{M}_1 \hat{L} f({\bf R},t),
        \label{Psi1formal}
        \\
        \Psi_2 ({\bf R},t) = (\hat{L} \hat{M}_1 \hat{L} \hat{M}_1 + \hat{L} \hat{M}_2) \hat{L} f({\bf R},t).
        \label{Psi2formal}
    \end{eqnarray}
\end{subequations}
Returning to Eq.\eqref{Operators} we can see that operator \eqref{LOperator} is diagonal in the plane wave basis and has the eigenvalue:
\begin{gather}\label{Leigenvalue}
 L_{{\bf q}\omega}=\frac{1}{\varepsilon_{{\bf q}}-i\gamma},
\end{gather}
where
\begin{gather}\label{dispers}
 \varepsilon_{{\bf q}} = \alpha T_c [\epsilon+\xi^2{\bf q}^2].
\end{gather}
So, it is convenient to deal with Fourier transformed functions, $\Psi ({\bf R},t) = \sum_{{\bf q}\omega} \Psi_{{\bf q}\omega} e^{i({{\bf qR}-\omega t})}$ and $f ({\bf R},t) = \sum_{{\bf q}\omega} f_{{\bf q}\omega}e^{i({{\bf qR}-\omega t})}$. Substituting \eqref{Psiformal} to \eqref{curterms2}, performing Fourier transformation and assuming $\gamma''\ll\gamma'$, after some computations, we arrive at expressions:
%
\begin{widetext}
\begin{subequations}\label{CurCom}
    \begin{eqnarray}
        \nonumber
        \langle {\bf j}^D\rangle  =
        -8e^3T (\alpha T_c \xi^2)^2 \sum\limits_{\textbf{p}} \frac{1}{\varepsilon_{-}}
        \Biggl\{
         \frac{\textrm{Re}[\textbf{A}(\textbf{p}\textbf{A}^*)]\gamma''\Omega}{(\varepsilon_{-}+\varepsilon_{+})^2 + \gamma'^2\Omega^2}
            \biggl[
                1 + \frac{2(\varepsilon_{-}-\varepsilon_{+})(\varepsilon_{-}+\varepsilon_{+})}{(\varepsilon_{-}+\varepsilon_{+})^2 + \gamma'^2\Omega^2}
            \biggr]
         +
        \\
        + \frac{2\textrm{Im}[\textbf{A}(\textbf{p}\textbf{A}^*)]\gamma'\gamma''\Omega^2(\varepsilon_{-}-\varepsilon_{+})} {[(\varepsilon_{-}+\varepsilon_{+})^2 + \gamma'^2\Omega^2]^2}
        \Biggr\}
        \label{CurDCom}
        \\\nonumber
        \langle {\bf j}^P\rangle = 8e^3T(\alpha T_c \xi^2)^3  \sum\limits_{\textbf{p}}
            \Biggl\{
                \frac{(\textbf{p}-\textbf{k})|\textbf{p}\textbf{A}|^2}{\varepsilon_-^2}
                \frac{\gamma''\Omega}{(\varepsilon_{-}+\varepsilon_{+})^2 + \gamma'^2\Omega^2}
                \biggl[
                    1 + \frac{(\varepsilon_{-}-\varepsilon_{+})(\varepsilon_{-}+\varepsilon_{+})} {(\varepsilon_{-}+\varepsilon_{+})^2 + \gamma'^2\Omega^2}
                \biggr] +
                \\
                + \frac{(\textbf{p}+\textbf{k})|\textbf{p}\textbf{A}|^2}{\varepsilon_-\varepsilon_+}              \frac{\gamma''\Omega(\varepsilon_{-}-\varepsilon_{+})(\varepsilon_{-}+\varepsilon_{+})} {[(\varepsilon_{-}+\varepsilon_{+})^2 + \gamma'^2\Omega^2]^2}
            \Biggr\}
        \label{CurPCom}
    \end{eqnarray}
\end{subequations}
\end{widetext}
where $\varepsilon_{\pm}=\varepsilon_{({\bf p}\pm{\bf k})/2}$ and ${\bf A}$ is a complex amplitude of vector potential ${\bf A}({\bf R},t) =
    {\bf A} e^{i({\bf kR}-\Omega t)} + c.c.$.
The full integration of expressions \eqref{CurCom} is quite difficult but the polar angle integration can be performed.
To make the text be not overloaded, we set the cumbersome integrals to the appendix section and produce the second-order conductivity in the following form:
\begin{subequations}\label{cond}
    \begin{eqnarray}
        \sigma^s_{\alpha\beta\gamma}(\tilde{T},\tilde{\Omega},\Theta) = \frac{\gamma''}{\gamma'}\frac{e^3\tilde{T}\xi^2I^s_{\alpha\beta\gamma}(\tilde{T},\tilde{\Omega},\Theta)} {\hbar^2 c\tilde{T}_c \tilde{\Omega}^4\cos^3(\Theta)},
            \label{Lin}
        \\
        \sigma^a_{yxy}(\tilde{T},\tilde{\Omega},\Theta) = - \frac{\pi \gamma''}{2\gamma'}\frac{e^3\tilde{T}\xi^2I^a_{yxy}(\tilde{T},\tilde{\Omega},\Theta)}{\hbar^2 c\tilde{T}_c^2 \tilde{\Omega}^5\cos^5(\Theta)},
        \label{Circ}
    \end{eqnarray}
\end{subequations}
where dimensionless factors $I^s_{\alpha\beta\gamma}$ and $I^a_{yxy}$ are given in the appendix, $\tilde{\Omega}=\Omega\xi/c$, $\tilde{T}_{(c)}=k_BT_{(c)}\xi/\hbar c$ and the relation $|\textbf{k}|=\cos(\Theta)\Omega/c$ has been used.

\section{Results and discussion}

\begin{figure}
 \includegraphics[width=\linewidth]{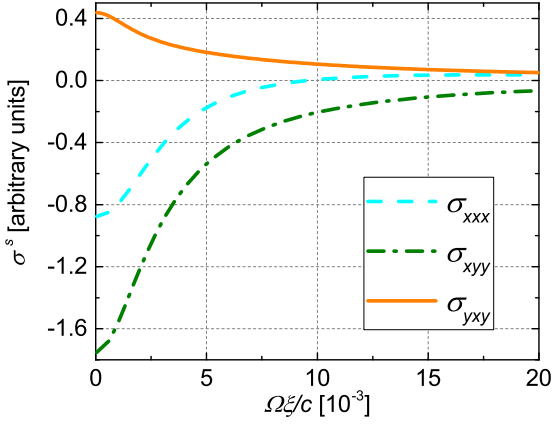}
 \caption{\label{fig2} The frequency dependence of symmetric part of the second-order conductivity calculated by Eq.\eqref{Lin}.}
\end{figure}

The qualitative dependence of \eqref{Lin} on dimensionless frequency $\tilde{\Omega}$ is shown in Fig.\ref{fig2}. It is proved that the absolute value of each component of the symmetrical part of the second-order conductivity monotonically increases, while the frequency decreases and, furthermore, it converges to the constant value at $\Omega=0$. The numerical computation shows that, in fact, no all components are independent and the following  equality is obeyed:
\begin{gather}\label{sym}
 \sigma^s_{xxx} - \sigma^s_{xyy} = 2\sigma^s_{yxy},
\end{gather}
where we omit arguments  $(\tilde{T},\tilde{\Omega},\Theta)$. This relation is not accidental and is a result of the system symmetry with regard to the rotation around the z-axis.
Usually, the wave-vector of EM-wave is the smallest in comparison with a wave-vector of any excitation in a solid. With this assumption,
we would expand the second-order conductivity in powers of ${\bf k}$: $\sigma_{\alpha\beta\gamma}({\bf k})\simeq \sigma_{\alpha\beta\gamma}(0) + D_{\alpha\delta\beta\gamma}k_\delta$, where $D_{\alpha\delta\beta\gamma}$ is a forth-rank tensor. Owing to the presence of the inversion symmetry in the system, the first term of expansion vanishes, $\sigma_{\alpha\beta\gamma}(0)=0$.
%
Then the requirement of invariance under the rotation at an arbitrary angle around the z-axis produces the relation $D_{xxxx}-D_{xxyy}=2D_{yxxy}$ that gives the formula \eqref{sym}.
But in deriving \eqref{CurCom}, the smallness of ${\bf k}$ is not used explicitly and that is clear from the dependence of $\varepsilon_\pm$ on $k$. However, we  can still represent the second-order conductivity in the form $\sigma_{\alpha\beta\gamma}({\bf k})= D_{\alpha\delta\beta\gamma}k_\delta$, which will be rotational-invariant. For example, let us consider the first term in \eqref{CurPCom} and rewrite it in the form:
\begin{gather}\nonumber
    j_\alpha^{P,1} =
    \sum_{\bf p}f(\varepsilon_\pm)(\textbf{p}-\textbf{k})|\textbf{p}\textbf{A}|^2 =
    \\
    \sum_{\bf p}f(\varepsilon_\pm)M_{\alpha\delta}p_\beta p_\gamma k_\delta A_\beta A_\gamma^* = D(|{\bf k}|)_{\alpha\delta\beta\gamma}k_\delta A_\beta A_\gamma^*,\label{rot}
\end{gather}
where $M_{\alpha\delta}$ is the matrix transforming ${\bf k}$ to $\textbf{p}-\textbf{k}$. Further it is not difficult to check that \eqref{rot} does not change under the rotation in the (x,y)-plane.
In practice, the light polarization is often defined by Stokes parameters. So, it is convenient to rewrite the first line of  \eqref{curmain} in the corresponding form:
\begin{gather}\label{Stokes}
 j_x = \frac{\sigma_{xxx}+\sigma_{xyy}}{2}(|E_x|^2+|E_y|^2) + \sigma_{yxy}(|E_x|^2-|E_y|^2).
\end{gather}

Further, the dependence of drag current magnitude on temperature arouses great interest. But, to begin with, it is necessary to confine the temperature range of applicability of the theory. First, the inequality $\epsilon\approx (T-T_c)/T_c\ll 1$ should be obeyed because the GL free energy \eqref{GLfEnergy} is derived under this condition. Second, the presented theory does not include the effects of interaction between superconductive fluctuations because we omit the term $\sim|\Psi|^4$ in the GL free energy. At an essentially small $\epsilon$ the fluctuations become strong and this contribution cannot be neglected. The analysis \cite{Larkin2005} produces the so-called Ginsburg-Levanyuk parameter $Gi\approx T_c/E_F$ which characterizes the temperature range of strong fluctuations. Our theory is correct for the case of weak fluctuations only, i.e. under the condition $\epsilon\gg Gi$.

\begin{figure}
 \includegraphics[width=\linewidth]{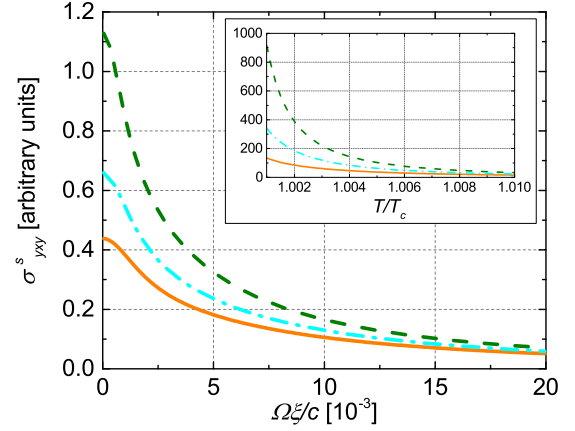}
 \caption{\label{fig3}The frequency dependence of $\sigma_{yxy}^s$ at different temperatures: $T/T_c=1.06$, $1.08$, $1.1$ for orange, blue and olive curves respectively. Inset: the temperature dependence of the same at different frequency: $\tilde{\Omega}=5\cdot10^{-3}$, $2.5\cdot10^{-3}$, $0.1\cdot10^{-3}$ for orange, blue and olive curves, respectively.}
\end{figure}

The temperature dependence of symmetrical part of the second-order conductivity is similar for each component. So, it is enough to show the qualitative results for one component only, for instance, for the $\sigma_{yxy}^s$-component (Fig.\ref{fig3}). It is proved that the current substantially increases when the temperature is close to its critical value.
To obtain the obvious temperature dependence, let us consider the range of small frequency.
For that purpose, the frequency should be turned to zero and that allows us to perform the integration in \eqref{Iyxy} explicitly. After some computations we obtain the simple expression:
\begin{gather}\label{ZeroFreq}
    \sigma_{yxy}^s(\Omega\rightarrow 0) = \frac{\gamma''}{\gamma'}\frac{e^3T\xi^2\cos(\Theta)}{48\hbar^2cT_c\epsilon^2}.
\end{gather}
We can see that the reduced temperature dependence at zero frequency, $\sigma^s \sim 1/\epsilon^2$, is rather dramatic and includes an additional power of  $\epsilon$ in comparison with the Aslamazov-Larkin conductivity.

\begin{figure}
 \includegraphics[width=\linewidth]{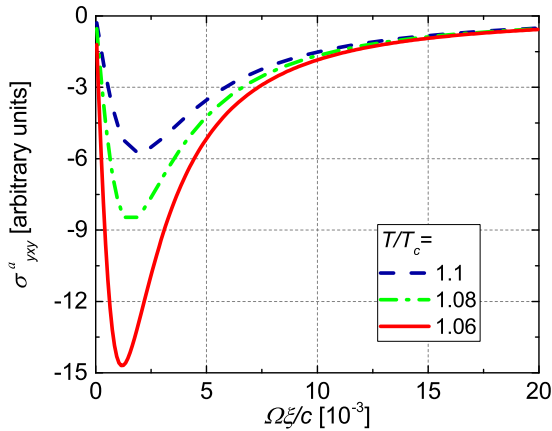}
 \caption{\label{fig4}The frequency dependence of antisymmetric part of the second-order conductivity calculated by Eq.\eqref{Circ}}
\end{figure}

The frequency dependence of \eqref{Circ} is non-monotonic and possesses its extremum at a small value of $\tilde{\Omega}$ (Fig.\ref{fig4}). With the decreasing temperature the extremum depth increases and moves towards zero frequency.
We want to remind here that the current, defined by the asymmetric component of conductivity, is nonzero in response to the circular-polarized EM-wave only, which is characterized by the direction of vector ${\bf E}$ rotation. Thus, the switching from the clockwise polarization to the reverse one changes the sign of $E_xE_y^*-E_x^*E_y$ as well as the direction of current $y$-projection.

The important feature of the obtained frequency dependence of the second-order conductivity consists in that its qualitative behaviour is the same as for the case of photon drag effect in conventional systems, for example, based on graphene \cite{Karch2010,Karch2010arXiv}.  Apparently, the reason of such similarity lies in a certain affinity of the TDGL-equation and the Boltzmann one, which is widely used for analyzing the nonlinear response of 2D electron gas.

At the end of this subsection, we discuss the magnitude of the examined effect.
For this purpose, let us compare the contribution of superconducting fluctuations with the one of normal electron gas. For the estimation, it is enough to use the simplest classical expression for the photon drag current in 2D systems, which has the following form \cite{Ivchenko2012}:
\begin{gather}\label{Ivch}
    {\bf j}_n = \frac{2e^3n}{\Omega m^2}\frac{\tau^2}{1+(\Omega\tau)^2} |E|^2 {\bf k},
\end{gather}
where $n$ is the electron gas density and $\tau$ is the momentum relaxation time.
With utilizing Eqs.\eqref{ZeroFreq} and \eqref{Ivch}, the ratio of two contributions in the zero-frequency limit can be easily composed:
\begin{gather}\label{Ratio}
 \frac{j^s}{j_n} = \frac{8}{3}\frac{\gamma''}{\gamma'}\frac{Tn\xi^2}{T_c}\left( \frac{\sigma_{AL}}{\sigma_n} \right)^2,
\end{gather}
where $\sigma_n=(e^2/h)k_Fl$ is the Drude conductivity and $\sigma_{AL}={e^2}/{16\hbar\epsilon}$ is the Aslamazov-Larkin conductivity for the 2D system.
Eq.\eqref{Ratio} is convenient to be considered piecemeal.
First, the ratio $T/T_c\approx 1$ and it does not play any role.
Second, to estimate the ratio $(\sigma_{AL}/\sigma_n)^2$, we take the typical values of normal conductivity $\sigma_n=10^{-3}\div 10^{-2}\,\Omega^{-1}$, that gives $(\sigma_{AL}/\sigma_n)^2\simeq 10^{-4}\div 10^{-2}$ at $\epsilon=0.1$.
Further, the quantity $|\gamma''/\gamma'|$ was supposed to be much less then unity. For instance, in tantalum nitride thin films, it takes the value $\sim 10^{-3}$ \cite{Breznay2012}. For TMD superconductors, the measurements of $|\gamma''/\gamma'|$ have not been performed yet. However, we can estimate it assuming the power dependence of $T_c$ on $E_F$, that gives $|\gamma''/\gamma'|\sim|\partial T_c/\partial E_F| \sim T_c/E_F\sim 10^{-3}$ at $T_c=9$ K and $n=10^{14}$ cm$^{-2}$ for MoS$_2$.
The above arguments give an idea that the ratio \eqref{Ratio} possesses a small magnitude. However, the rest dimensionless factor $n\xi^2$ is often very large.
In TaN films $n\xi^2\simeq 3\cdot 10^{4}$, while, in MoS$_2$, $n\xi^2\approx n\xi_0 l\simeq 5\cdot 10^{3}$ (where $\xi_0$ is the BCS coherent length at $T=0$ K and $l$ is mean free path).
Finally, the estimation becomes $j^s/j_n\sim 10^{-3}\div 1$ and points to the possibility of experimental observation of this effect.



\section{Conclusion}

In the presented work we developed the theory of photon drag of the superconducting fluctuations based on using the TDGL-equation.
The calculation showed that the magnitude of the photo-drag current strongly grows when the temperature comes down to the critical point.
In the low-frequency domain the drag-current is proportional to the squared Aslamazov-Larkin conductivity, that was not evident from the beginning.
We would like to emphasize that
the induced current is proportional to the imaginary part of the $\gamma$-parameter as it has a place in the Hall-effect  and, consequently, the presented effect can be treated as an additional approach in the fluctuation spectroscopy.
It is interesting to note that in thin superconducting films with three-dimensional electrons
and a simple electron spectrum,  parameter $\gamma''$ is negative. Thus, the photon drag of fluctuations will compensate the photocurrent of normal electrons. Taking into account the commensurability of these currents, as it was shown by the estimation above, the reduction of the full photocurrent near $T_c$ can be considerable.

\begin{acknowledgments}
We thank A.G. Semenov for his useful discussion.
This work was supported by the Foundation for the Advancement of Theoretical Physics and Mathematics "BASIS", RFBR (Grant No. 18-29-20033) and by the Ministry of Science and Higher Education of the Russian Federation (project "Nonlinear electrodynamics of electron systems in micro- and nanostructures").
\end{acknowledgments}

\appendix*
\section{Explicit expressions for integrals $I_{\alpha\beta\gamma}$}

Let us introduce for brevity the following notations:

\begin{gather}
    a=a(T,\tilde{\Omega},\Theta)=\frac{4\epsilon}{\tilde{\Omega}^2\cos^2(\Theta)}
    \\
    b=b(\tilde{\Omega},\Theta)=\left(\frac{\pi}{4\tilde{T}_c\tilde{\Omega}\cos^2(\Theta)}\right)^2
\end{gather}
%
Then the dimensionless factors from the resulting expressions \eqref{cond} have the  form:

\begin{widetext}

\begin{gather}
    I_{yxy}^a(T,\tilde{\Omega},\Theta) = \int\limits_{1+a}^{\infty}dy\,
    \frac{1}{\left(y^2+b\right)^2}
    \frac{y\left(y-\sqrt{(y-2)^2+4a}\right)}
    {\sqrt{(y-2)^2+4a}},
\end{gather}
\begin{gather}
    I_{xxx}^s(T,\tilde{\Omega},\Theta) = 2\int\limits_{1+a}^{\infty}dy\,
    \frac{1}{\left(y^2+b\right)^2}
    \left\{
        \left[
            y^2+b(2y-1)
        \right]
        \left[
            1-\frac{y}{\sqrt{(y-2)^2+4a}}
        \right]
        +\frac{4by(y-2)(y-1-a)}{[(y-2)^2+4a]^{3/2}}
    \right\},
\end{gather}
\begin{gather}
    I_{xyy}^s(T,\tilde{\Omega},\Theta) = -4\int\limits_{1+a}^{\infty}dy\,
    \frac{1}{\left(y^2+b\right)^2}
    \left\{
        \left[
            y^2+b(y-1)
        \right]
        \left[
            1-\frac{y}{\sqrt{(y-2)^2+4a}}
        \right]
        +\frac{2(y-1-a)(b+2y)}{\sqrt{(y-2)^2+4a}}
    \right\},
\end{gather}
\begin{gather}\label{Iyxy}
    I_{yxy}^s(T,\tilde{\Omega},\Theta) = -\int\limits_{1+a}^{\infty}dy\,
    \frac{1}{\left(y^2+b\right)^2}
    \left\{
        \left[
            y^2+b(4y-1)
        \right]
        \left[
            1-\frac{y}{\sqrt{(y-2)^2+4a}}
        \right]
        +\frac{8b(y-1-a)}{\sqrt{(y-2)^2+4a}}
    \right\},
\end{gather}

\end{widetext}

\bibliography{bib}

\end{document}